\documentclass[11pt]{article}
\usepackage{amsmath}
\usepackage{tensor}
\usepackage{amsfonts}
\usepackage[hmargin={25mm,25mm},vmargin={23mm,24mm}]{geometry}
\usepackage{hyperref}
\usepackage{verbatim}
\usepackage[shortlabels]{enumitem}
\usepackage{slashed}
\usepackage{braket}
\usepackage{caption}
\usepackage{cite}
\usepackage{nicefrac}
\usepackage{bbm}
\usepackage{url}
\usepackage{dsfont}

\numberwithin{equation}{section}

\def \ifempty#1{\def\temp{#1} \ifx\temp\empty }

\newcommand{\bareps}{\bar{\epsilon}}
\newcommand{\deltahat}{\hat{\delta}}
\newcommand{\etabar}{\bar{\eta}}

\newcommand{\BetaijBeta}[2]{\left( \bar{\beta}\Gamma_- \Gamma_{#1 #2} \beta \right)}

\newcommand{\BetaIijBeta}[3]{\left( \bar{\beta}\Gamma_- \Gamma^{#1} \Gamma_{#2 #3} \beta \right)}

\newcommand{\EpsIBeta}[1]{\left( \bar{\epsilon}_+ \Gamma_- \Gamma^{#1} \beta \right)}

\newcommand{\EpsijBeta}[2]{\left( \bar{\epsilon}_+ \Gamma_- \Gamma_{#1 #2} \beta \right)}
\newcommand{\EpsIijBeta}[3]{\left( \bar{\epsilon}_+\Gamma_- \Gamma^{#1} \Gamma_{#2 #3} \beta \right)}
\newcommand{\EpsIJijBeta}[4]{\left( \bar{\epsilon}_+\Gamma_- \Gamma^{#1 #2} \Gamma_{#3 #4} \beta \right)}

\begin{document}
\vspace*{0mm}
\begin{center}
\LARGE \bf {\sc  Supersymmetric Soliton $\sigma$-Models from Non-Lorentzian Field Theories}
\end{center}

\centerline{\LARGE \bf {\sc  }} \vspace{2truecm} \thispagestyle{empty} \centerline{
    {\large {\bf {\sc Rishi Mouland }}}\footnote{E-mail address: \href{mailto:rishi.mouland@kcl.ac.uk}{\tt rishi.mouland@kcl.ac.uk}}  }

\vspace{1cm}
\centerline{ {\it Department of Mathematics}}
\centerline{{\it King's College London, }} 
\centerline{{\it  WC2R 2LS, UK}} 

\vspace{1.0truecm}

 
\thispagestyle{empty}

\centerline{\sc Abstract}
\vspace{0.4truecm}
\begin{center}
\begin{minipage}[c]{360pt}{
    }
Recently, a class of non-Lorentzian supersymmetric Lagrangian field theories was considered, in some cases describing M-theory brane configurations, but more generally found as fixed points of non-Lorentzian RG flows induced upon Lorentzian theories. In this paper, we demonstrate how the dynamics of such theories can be reduced to motion on the supersymmetric moduli space of BPS solitons of the parent theory. We focus first on the $\mathcal{N}=(1,1)$ $\sigma$-model in $(1+1)$-dimensions with potential, where we produce a supersymmetric extension to the standard geodesic approximation for slow kink motion. We then revisit the $(4+1)$-dimensional Yang-Mills-like theory with 24 supercharges describing a null compactification of M5-branes. We show that the theory reduces to a $\sigma$-model on instanton moduli space, extended by couplings to additional fields from the parent theory, and possessing (8+8) super(conformal) symmetries along with 8 further fermionic shift symmetries. We derive this model explicitly for the single $SU(2)$ instanton.

\end{minipage}
\end{center}

\newpage 
 
\renewcommand{\thefootnote}{\arabic{footnote}}
\setcounter{footnote}{0} 

\section*{Introduction}

The study of String Theory and M-Theory relies on our understanding of the dynamics of strings and branes, which has in turn motivated the longstanding study of supersymmetric field theories. While one usually considers Lorentz-invariant theories, there are many scenarios in which non-Lorentzian theories can arise, such as when a fixed frame of reference or Lorentz-violating background are chosen. In many cases, the result is a field theory with symmetry group given by some contraction of the Lorentz group, such as the Galilean or Carrollian symmetry groups \cite{Duval:2014uoa,Leblanc:1992wu,Bandos:2008fr,Bagchi:2015qcw}. Also of interest particularly in condensed matter systems are theories with Lifshitz scaling symmetry \cite{Orlando:2009az,Gomes:2014tua,Chapman:2015wha,Xue:2010ih}.

Another well established topic in high energy theory concerns classical vacua in possibly topologically non-trivial sectors of a theory's configuaration space: solitons. In particular, one is often interested in solitons that saturate some BPS bound, as in supersymmetric theories these describe non-perturbative vacua preserving some amount of supersymmetry. Central to the study of the low-energy scattering of solitons was the work of Manton \cite{Manton:1981mp} on BPS monopoles, where it was first argued that the dynamics of slowly moving solitons can be captured by geodesic motion on the moduli space of static solutions. Later work explored many different avenues, including applying similar methods to Yang-Mills solitons of other codimension \cite{Ruback:1988ba,Ward:1985ij}, and most relevant  to this work, the study of a supersymmetric extension to this approximation \cite{Gauntlett:1992yj,Cederwall:1995bc,Gauntlett:1995fu,Moss:1998jf} to determine the supersymmetric effective theory for both the bosonic and fermionic soliton zero modes.\\

In recent work \cite{Lambert:2019nti}, a procedure was described in which one induces a non-Lorentzian RG flow on a Lorentz-invariant supersymmetric Lagrangian field theory, described by a parameter $\eta$. While for $\eta\neq 0$ the flow is invertible and uninteresting, the Lagrangian diverges as we take $\eta\to 0$. However, a non-Lorentzian Lagrangian description for the fixed point theory was proposed, exhibiting preserved or even enhanced supersymmetry. The construction necessitates the introduction of a Lagrange multiplier field, due to which the dynamics reduces to motion on the moduli space of $\frac{1}{2}$-BPS solitons of the parent theory.

The purpose of this paper is to make explicit the reduction of such non-Lorentzian fixed point theories to supersymmetric quantum mechanics on various soliton moduli spaces. Such models will generically feature both dynamical coordinates on the target (moduli) space, as well as couplings to time-dependant parameters descended from additional fields in the parent theory. We will focus on two particular examples.

This paper is organised as follows. In section \ref{sec: general procedure}, we briefly review the construction of \cite{Lambert:2019nti} and outline the general procedure by which dynamics is reduced to quantum mechanics on soliton moduli space. In section \ref{sec: sigma models}, we present a particularly straightforward application of this procedure for the case of kinks in the $\mathcal{N}=(1,1)$, $(1+1)$-dimensional $\sigma$-model, where the procedure serves as a supersymmetric extension of the standard scaling argument for the geodesic approximation \cite{Manton:1981mp}. In section \ref{sec: SYM}, we revisit the non-Lorentzian $(4+1)$-dimensional $\mathcal{N}=2$ Yang-Mills-like theory first described in \cite{Lambert:2010wm,Lambert:2018lgt} as a non-Abelian M5-brane worldvolume theory, and later obtained via a rescaling of regular 5d super-Yang-Mills in \cite{Lambert:2019nti}. This theory has also more recently been recovered as a special case of a more general DLCQ prescription for the M5-brane, via holography \cite{Lambert:2019jwi}. We show that the theory reduces to a $\sigma$-model on the moduli space of a general $k$-instanton in $SU(N)$, and that in doing so all supersymmetry is preserved. The reduction is performed explicitly for the case of gauge group $SU(2)$ and instanton number $k=1$, or equivalently, a single unit of momentum on the compact, null M-Theory circle as per the DLCQ description of the M5-brane \cite{Aharony:1997th,Aharony:1997an}. In particular, the QM model takes the form of an $\mathcal{N}=(4,4)$ $\sigma$-model on the 8-dimensional moduli space $\mathcal{M}_{2,1}=\mathbb{R}^4\times \left( \mathbb{R}^4/\mathbb{Z}_2 \right)$ of a single $SU(2)$ instanton, coupled to time-dependant parameters describing the finite energy zero modes of the initial worldvolume theory's embedding coordinates $X^I$ and electric potential $A_0$. Finally, in  Section \ref{sec: conclusion} we summarise our conclusions.

\section{Non-Lorentzian RG-flows and their fixed point actions}\label{sec: general procedure}

We first briefly review the construction of \cite{Lambert:2019nti} for a general supersymmetric Lagrangian field theory. A rescaling of the coordinates and fields was considered with real parameter $\eta$, under which the action and supersymmetry variation split as
\begin{align}
  S &= \eta^{-1} S_{-1} + S_0 + \eta S_1 +\dots \nonumber\\
  \delta &= \eta^{-1} \delta_{-1} + \delta_0 + \eta \delta_1 +\dots\ ,
\end{align}
with $S_{-1}$ schematically taking the form
\begin{align}
  S_{-1} = \int d^d x\,\, \Omega^2\ ,
\end{align}
for some $\Omega$ made up of bosonic fields. In the limit $\eta\to 0$, dynamics are localised to the surface in configuration space defined by $\Omega=0$. It is shown that one can always choose $\tilde{\delta}$ such that
\begin{align}
  ( \delta_0 + \tilde{\delta} ) ( S_0 + \tilde{S} ) =0\ ,
\end{align}
where
\begin{align}
  \tilde{S} = \int d^d x\,\, G\, \Omega
\end{align}
introduces a Lagrange multiplier to impose $\Omega=0$, and the shift to the supersymmetry variation $\tilde{\delta}$ generically acts on both the original fields of the theory, and $G$. Then, $S_0 + \tilde{S}$ is proposed as a Lagrangian description of the theory at the fixed point $\eta\to 0$, which by construction enjoys a Lifshitz scaling symmetry (with suitable scaling dimension chosen for $G$), as well as the full supersymmetry of the parent theory.

This procedure is performed explicitly in a variety of cases, all relevant to M-theory branes: maximally supersymmetric ($\mathcal{N}=2$) super-Yang-Mills in five dimensions, the scaling limit of which can be interpreted as recovering the DLCQ description of a stack of M5-branes (for gauge group $U(N)$); and both the $\mathcal{N}=6$ (ABJM) and $\mathcal{N}=8$ (BLG) Chern-Simons-matter theories describing stacks of M2-branes, the scaling limit of which describes the U-dual of the M5-brane DLCQ. Each of these cases highlighted interesting aspects of the procedure. In the former case, the supersymmetry was enhanced from 16 to 24 super(conformal) symmetries, a phenomenon that was later elucidated by a holographic interpretation of the scaling limit \cite{Lambert:2019jwi}, while in the latter case, a Lagrangian description of the fixed point theory required the theory's bosonic field content to be mixed and reformed into new fields behaving homogeneously under the RG flow.\\

The remainder of this paper is concerned with taking the next step; namely, integrating out the Lagrange multiplier $G$ as well as other non-dynamical fields and thus reducing the dynamics to a constraint surface $\Sigma$ in configuration space. In the cases considered, a common feature is that as well as having $G$ imposing $\Omega=0$, which for these examples will be a Bogomol'nyi-like equation defining a topological soliton, we also find a fermionic Lagrange multiplier descended from one of the two chiral components of a fermion of the parent theory, which imposes a Dirac equation on the other component. For the case of super-Yang-Mills, we will have additional non-dynamical fields which further constrain $\Sigma$.

Then, to determine the constrained theory, we solve the constraints defining $\Sigma$ and evaluate the action on this solution. One can then use the supersymmetry of the parent theory to determine how the coordinates on $\Sigma$ transform under supersymmetry in the reduced theory. In considering this procedure for the M5-brane theory in section \ref{sec: SYM}, the presence of additional constraints requires some care, in particular in ensuring that supersymmetry is preserved on $\Sigma$.

\section{Kinks in the $\mathcal{N}=(1,1)$, $(1+1)$-dimensional $\sigma$-model with potential}\label{sec: sigma models}

We first present a scaling limit and the subsequent reduction to a quantum mechanical model for perhaps the simplest supersymmetric theory containing BPS solitons: the $\mathcal{N}=(1,1)$, $(1+1)$-dimensional $\sigma$-model.

\subsection{The parent theory}

We consider the $(1+1)$-dimensional $\sigma$-model with $d$-dimensional Riemannian target manifold $(M,g)$, local coordinates $\phi^i$ and generic scalar potential $W(\phi)$. Letting $\Xi$ denote the 2-dimensional worldsheet with signature $(-,+)$, we have action
  \begin{align}
  S = \frac{1}{2} \int_\Xi d^2 x\,\, \left( -g_{ij} \partial_\mu \phi^i \partial^\mu \phi^j - g^{ij} D_i W D_j W + i g_{ij} \bar{\psi}^i \slashed{D}\psi^j - \frac{1}{6} R_{ijkl} \bar{\psi}^i \psi^k \bar{\psi}^j \psi^l + i \left( D_i D_j W \right)\bar{\psi}^i \psi^j \right)\ .
\end{align}
The $\left\{ \psi^i \right\}_{i=1}^d$ are two-component Majorana spinors, with conjugate $\bar{\psi}^i= (\psi^i)^T \gamma_t$. The covariant derivative of the $\psi^i$ is defined in terms of the pullback of the Levi-Civita connection $\Gamma^i_{jk}$ in coordinate basis $\{\phi^i\}$ on $M$,
\begin{align}
  D_\mu \psi^j = \partial_\mu \psi^j + \Gamma^j_{kl}(\partial_\mu \phi^k)\psi^l\ .
\end{align}
We then find that $S$ enjoys $\mathcal{N}=(1,1)$ supersymmetry,
\begin{align}
  \delta \phi^i &= i\bar{\epsilon} \psi^i \nonumber\\
  \delta \psi^i &= -\slashed{\partial} \phi^i \epsilon + (D^i W)\epsilon - i \Gamma^i_{jk} (\bareps \psi^j)\psi^k \ .
\end{align}
Note, in general one can consider further potential terms for $S$ in terms of Killing vector fields on $M$ \cite{AlvarezGaume:1983ab}, which we choose to omit.\\

Suppose now that $D_i W$ has zeros at points $p\in \mathcal{P}\subset M$, and is non-zero elsewhere. Thus, any finite energy configuration $\phi$ must satisfy $\lim_{x\to \pm\infty}\phi\in \mathcal{P}$. Further imposing that the points in $\mathcal{P}$ are isolated, we see that the space of finite energy configuration splits into topological sectors corresponding to which element of $\mathcal{P}$ $\phi$ settles at as $x\to \pm \infty$.

We briefly review the construction of classical \textit{static} solutions in each topological sector. The bosonic part of the energy functional is given by
\begin{align}
  E_\text{bos.} = \frac{1}{2} \int dx \left( g_{ij} \partial_t\phi^i \partial_t\phi^j + g_{ij} \partial_x\phi^i \partial_x\phi^j + g^{ij} D_i W D_j W \right)\ .
\end{align}
So, setting $\partial_t\phi^i=0$, we perform the standard Bogomol'nyi recasting of $E_\text{bos.}$ to arrive at
\begin{align}
  E_\text{bos.} = \frac{1}{2} \int dx &\,\, g_{ij} \left( \partial_x\phi^i \pm D^i W \right)\left( \partial_x\phi^j \pm D^j W \right) \nonumber\\
  &\mp \left( W(\phi(\infty)) - W(\phi(-\infty)) \right)\ .
  \label{eq: Bogomol'nyi argument}
\end{align}
Hence, we have
\begin{align}
  E_\text{bos.} \ge |W(\phi(\infty)) - W(\phi(-\infty))|\ ,
\end{align}
with equality precisely for configurations satisfying
\begin{align}
  \left\{\begin{aligned}
  \partial_x\phi^i - D^i W = 0 \quad &\text{if } W(\phi(\infty)) - W(\phi(-\infty))\ge 0\quad\text{(kink)}\\
  \partial_x\phi^i + D^i W = 0 \quad &\text{if } W(\phi(\infty)) - W(\phi(-\infty))\le 0\quad\text{(anti-kink)}
  \end{aligned}\right.
  \label{eq: Bogomol'nyi equations}
\end{align}
It is easily verified that such kink solutions satisfy the second-order bosonic equations of motion for $\phi^i$, and are thus classical bosonic solutions. Upon solving (\ref{eq: Bogomol'nyi equations}) for the relevant topological sector, one can then use supersymmetry to generate fermionic zero modes.

These static bosonic solutions define $\tfrac{1}{2}$-BPS states of the theory. This is easily seen by decomposing spinors as $\psi^i = \psi^i_+ + \psi^i_-$, with $\psi^i_\pm = \frac{1}{2}\left( 1\pm \gamma_x \right)\psi^i$ and similarly for $\epsilon$, and noting
\begin{align}
  \delta\psi^i_+ &= \gamma_t \partial_t\phi^i \epsilon_- - \left( \partial_x\phi^i - D^i W \right) \epsilon_+ + \text{fermions} \nonumber\\
  \delta\psi^i_- &= \gamma_t \partial_t\phi^i \epsilon_+ + \left( \partial_x\phi^i + D^i W \right) \epsilon_- + \text{fermions}\ ,
\end{align}
and so depending on the sign of $W(\phi(\infty)) - W(\phi(-\infty))$, one of $\epsilon_\pm$ is broken, while the other preserved\footnote{Of course, if $W(\infty)-W(-\infty)=0$ then we have the trivial vacuum $\phi=\text{constant}\in\mathcal{P}$, and full supersymmetry is preserved.}.

\subsection{Scaling limit and fixed point action}

We now consider a rescaling of the fields and worldsheet coordinates of the theory with parameter $\eta$. In particular, we choose our scaling such that the small $\eta$ limit corresponds to slow motion of the coordinates $\phi^i$, i.e. $\partial_t \phi^i << \partial_x \phi^i$, and hence the theory at the fixed point $\eta\to 0$ will provide an effective theory for the leading order dynamics of a slow moving (i.e. low energy) kink.

We expect to recover the standard argument after Manton \cite{Manton:1981mp} that these dynamics are described by geodesic motion on the moduli space of the static solution. In this construction, the metric on the moduli space descends from the kinetic terms in the action. To be more precise, if $\delta_\alpha \phi^i$ denotes the variation of $\phi^i$ with respect to a modulus $m^\alpha$, then this metric is
\begin{align}
  \mathcal{G}_{\alpha \beta} = \int dx\,\, g_{ij} \delta_\alpha \phi^i \delta_\beta \phi^j\ .
  \label{eq: induced metric}
\end{align}
One can propose an analogous metric for the Grassmann moduli \cite{Gauntlett:1992yj}, and thus try to construct a supersymmetric quantum mechanical model that pairs the bosonic and fermionic perturbations around the static bosonic kink. We will see that precisely this form for the quantum mechanics emerges as we take $\eta\to 0$.\\

Without loss of generality, we seek an effective action describing the slow motion of a kink rather than anti-kink, and so restrict $\phi^i$ to lie in the sectors of configuration space satisfying $W(\phi(\infty)) - W(\phi(-\infty))\ge 0$. The vacua in such sectors preserve the supersymmetry $\epsilon_+$, which pairs $\phi^i$ with $\psi^i_-$. Any fluctuations above these vacua will fall into representations of this unbroken supersymmetry $\epsilon_+$.

So we seek a scaling of both the coordinates and fields such that the limit $\eta\to 0$ will provide us with an effective theory for the leading order dynamics about a static kink solution. Such a scaling should satisfy the following.
\begin{itemize}
	\item Velocities should be suppressed relative to spatial variation, i.e. $\partial_t << \partial_x$
	\item We want to describe the dynamics of the perturbations about the static solution, and so the kinetic part of the action for $\phi^i$ should appear in the scaled theory at order $\eta^0$
	\item The original static kink should remain a classical solution, implying that $\left( \partial_x \phi^i - D^i W \right)$ should scale homogeneously
	\item As we take $\eta\to 0$, $\phi^i$ should remain paired to $\psi^i_-$ under the supersymmetry $\epsilon_+$
\end{itemize}
These requirements determine an essentially unique choice of scaling, given by 
\begin{align}
  t &\to \eta^{-1} t & \psi^i_+ &\to \eta^{3/4} \psi^i_+ \nonumber\\
  x &\to \eta^{-1/2} x & \psi^i_- &\to \eta^{1/4} \psi^i_- \nonumber\\
  \phi^i &\to \eta^{-1/4} \phi^i & \epsilon_+ &\to \eta^{-1/2} \epsilon_+ \nonumber\\
  && \epsilon_- &\to  \epsilon_-\ ,
\end{align}
where we use the two degrees of scaling symmetry already present in $S$ (scaling by worldsheet `mass dimension', and local scaling diffeomorphisms of $M$) to keep $g_{ij}$ and $W$ fixed under the scaling by $\eta$. The action and supersymmetry variations then take the form
\begin{align}
  S &= \eta^{-1} S_{-1} + S_0 + \eta S_1 \nonumber\\
  \delta &= \eta^{-1} \delta_{-1} + \delta_0 + \eta \delta_1\ ,
\end{align}
with
\begin{align}
  S_{-1} &= \frac{1}{2} \int_\Xi d^2 x\,\, \Big( -g_{ij} \partial_x \phi^i \partial_x \phi^j - g^{ij} D_i W D_j W \Big) \nonumber \\
  S_0 &= \frac{1}{2} \int_\Xi d^2 x\,\, \left( g_{ij} \partial_t\phi^i \partial_t\phi^j - i g_{ij} \bar{\psi}^i_- \gamma_t D_t \psi^j_- - 2i \bar{\psi}^i_+\big( g_{ij} D_x - \left( D_i D_j W \right) \big)\psi^j_- \right) \nonumber\\
  S_1 &= \frac{1}{2} \int_\Xi d^2 x\,\, \left( -ig_{ij} \bar{\psi}^i_+ \gamma_t D_t \psi^j_+ -\tfrac{1}{3}\left( R_{ijkl}+R_{ilkj} \right)\bar{\psi}^i_+ \psi^k_- \bar{\psi}^j_+ \psi^l_-  \right)
  \label{eq: rescaled kink action}
\end{align}
\begin{align}
  \delta_{-1} \psi^i_+ &= - \left( \partial_x\phi^i - D^i W \right)\epsilon_+\nonumber\\[0.5em]
  \delta_0 \phi^i & = i\bareps_+ \psi^i_- \nonumber\\
  \delta_0 \psi^i_+ & = \gamma_t \partial_t\phi^i \epsilon_ - + i \Gamma^i_{jk}( \bar{\psi}^j_+ \psi^k_- ) \epsilon_+ \nonumber\\
  \delta_0 \psi^i_- & = \gamma_t \partial_t\phi^i \epsilon_+ + \left( \partial_x\phi^i + D^i W \right) \epsilon_-  \nonumber\\[0.5em]
  \delta_1 \phi^i & = i\bareps_- \psi^i_+ \nonumber\\
  \delta_1 \psi^i_- & =  i \Gamma^i_{jk}( \bar{\psi}^j_+ \psi^k_- ) \epsilon_-\ .
\end{align}
We've used that for a torsion-free connection, $\Gamma^i_{jk} ( \bareps_+ \psi^j_- ) \psi^k_- = \Gamma^i_{jk} ( \bareps_- \psi^j_+ ) \psi^k_+=0$.\\

In the $\eta\to 0$ limit, we localise onto classical minima of $S_{-1}$, i.e. solutions to $\partial_x\phi^i- D^i W=0$. By utilising a generalisation of the procedure of \cite{Lambert:2019nti} for non-constant metrics in the quadratic constraint imposed by $S_{-1}$, we propose the following action for the theory at the fixed point $\eta\to 0$,
\begin{align}
  \tilde{S} = \frac{1}{2} \int_\Xi d^2 x\,\, \Big(& g_{ij} \partial_t\phi^i \partial_t\phi^j + 2 G_i \left( \partial_x\phi^i - D^i W \right) \nonumber\\ 
  &- i g_{ij} \bar{\psi}^i_- \gamma_t D_t \psi^j_- - 2i \bar{\psi}^i_+\big( g_{ij} D_x - \left( D_i D_j W \right) \big)\psi^j_- \Big)\ ,
  \label{eq: kink fixed point action}
\end{align}
where here the target space 1-form $G_i(t,x)$ acts as a Lagrange multiplier imposing the Bogomol'nyi equation. $\tilde{S}$ then has $\mathcal{N}=(1,1)$ superysmmetry, given by
\begin{align}
  \tilde{\delta} \phi^i & = i\bareps_+ \psi^i_- \nonumber\\
  \tilde{\delta} \psi^i_+ & = \gamma_t \partial_t\phi^i \epsilon_- + i \Gamma^i_{jk}( \bar{\psi}^j_+ \psi^k_- ) \epsilon_+ +  G^i \epsilon_+ \nonumber\\
  \tilde{\delta} \psi^i_- & = \gamma_t \partial_t\phi^i \epsilon_+ + \left( \partial_x\phi^i + D^i W \right) \epsilon_- \nonumber\\
  \tilde{\delta} G_i &= i\bareps_+\left( \Gamma^j_{ik} G_j \psi^k_- + g_{ij}\gamma_t D_t \psi^i_+ + \tfrac{1}{3}\left( R_{ijkl}+R_{ilkj} \right)\psi^k_- (\bar{\psi}^j_+\psi^l_-) \right)\nonumber\\
  &\qquad + i\bareps_-\left( -g_{ij} D_x \psi^j_+ + (D_i D_j W)\psi^j_+ \right)\ .
  \label{eq: reduced sigma model SUSY}
\end{align}

\subsection{Reduction to constraint surface}\label{subsec: reduction to constraint surface}

The Lagrange multiplier $G_i$ imposes the constraint $\phi^i_x - D^i W=0$, and hence $\phi^i$ is constrained to live on the kink moduli space, where crucially the moduli $m^\alpha(t)$ are allowed to depend on time. Then, the bosonic part of the action on this constraint surface is given by
\begin{align}
  \tilde{S}_\Sigma^\text{bos.} = \frac{1}{2}\int_\Xi d^2 x\,\, g_{ij} \partial_t\phi^i \partial_t\phi^j \Big|_{\phi^i_x - D^i W=0} = \frac{1}{2} \int dt\,\, \mathcal{G}_{\alpha\beta} \dot{m}^\alpha \dot{m}^\beta\ ,
\end{align}
with $\mathcal{G}_{\alpha\beta}$ as defined in (\ref{eq: induced metric}). Thus, as expected we reproduce the standard moduli space approximation.

We also have a fermionic Lagrange multiplier, given by the now non-dynamical $\psi_+^i$. This imposes a fermionic constraint, such that the complete constraint surface can be denoted $\Sigma=\{\mathcal{C}_1^i=0,\mathcal{C}_2^i=0\}$, where
\begin{align}
  \mathcal{C}^i_1 &= \partial_x\phi^i - D^i W \nonumber\\
  \mathcal{C}^i_2 &= D_x \psi^i_- - (D^i D_j W) \psi^j_-\ .
  \label{eq: sigma model constraints}
\end{align}
We then find\footnote{We necessarily have that the full set of equations of motion from $\tilde{S}$ transform into one-another under $\tilde{\delta}$. We see here, however, that the constraints defining $\Sigma$ transform only amongst themselves}
\begin{align}
  \tilde{\delta}\mathcal{C}_1 &= i \bareps_+ \left( \mathcal{C}^i_2 - \Gamma^i_{jk} \mathcal{C}^j_1 \psi^k_- \right) \nonumber\\
  \tilde{\delta}\mathcal{C}_2 &= \left( D_t \mathcal{C}^i_1 + \tfrac{i}{2} R^i_{jkl} \mathcal{C}^i_1 ( \bar{\psi}^k_- \gamma_t \psi^l_- ) - i \Gamma^i_{jk} (\bar{\psi}^j_- \gamma_t \mathcal{C}^k_2) \right)\gamma_t \epsilon_+ \nonumber\\
  &\qquad +\left( D_x \mathcal{C}^i_1 + (D^i D_j W)\mathcal{C}^j_1 \right)\epsilon_-\ ,
\end{align}
and hence we are able to restrict to $\Sigma$ without breaking any supersymmetry. In principle, we can then solve the constraints (\ref{eq: sigma model constraints}) to determine $\phi^i$ and $\psi^i_-$ in terms of a set of both bosonic and fermionic time-dependant moduli. Upon substituting these forms for $\phi^i$ and $\psi^i_-$ back into $\tilde{S}$, we determine the supersymmetrised low energy effective action. The supersymmetry transformation rules for the moduli are determined from those of $\phi^i$ and $\psi^i_-$ (\ref{eq: reduced sigma model SUSY}). In particular, we see that the bosonic and fermionic moduli are paired under the supersymmetry $\epsilon_+$, while the `broken' supersymmetry $\epsilon_-$ lives on as a shift symmetry for its Goldstino mode $\psi^i_-$.\\

Finally, it is instructive to consider an explicit example. We take $(M,g)=(\mathbb{R},\delta)$ and $W(\phi)=\lambda\left( a^2 \phi - \tfrac{1}{3}\phi^3 \right)$. This choice describes a single scalar field in a `double dip' potential $(\partial_\phi W)^2=\lambda^2 \left( \phi^2 - a^2 \right)^2$. Finite energy configurations have $\lim_{x\to \pm\infty}\phi=\pm a$. Restricting to sectors satisfying $W(\phi(\infty)) - W(\phi(-\infty))\ge 0$ corresponds to requiring $\phi(\infty)\ge \phi(-\infty)$.

We now solve the constraints $\mathcal{C}_1^i=0,\mathcal{C}_2^i=0$ for the sector defined by $\phi(\infty)=a, \phi(-\infty)=-a$, i.e. the kink solution. These read
\begin{align}
  \partial_x\phi - \lambda \left( a^2 - \phi^2 \right) &=0 \nonumber\\
  \partial_x \psi_- + 2\lambda \phi \psi_- &= 0\ .
\end{align}
We find the general solutions
\begin{align}
  \phi(t,x) &= a \tanh \left[ \lambda a \left( x - y(t) \right) \right] \nonumber\\
  \psi_- (t,x) &= -\lambda a^2\,\text{sech}^2\left[ \lambda a \left( x-y(t) \right) \right] \chi_-(t)\ ,
\end{align}
where $y(t)\in\mathbb{R}$ is the position of the kink at time $t$, and $\chi_-(t)$ is a negative chirality spinor. The action $\tilde{S}$ evaluated on $\Sigma$ is then
\begin{align}
  \tilde{S}_\Sigma = \frac{2}{3} \lambda a^3 \int dt\,\, \Big( \left( \partial_t y \right)^2 - i \bar{\chi}_- \gamma_t \partial_t \chi_- \Big)\ ,
\end{align}
with supersymmetry
\begin{align}
  \tilde\delta y &= i \bareps_+ \chi_- \nonumber \\
  \tilde{\delta} \chi_- &= \left( \partial_t y \right) \gamma_t \epsilon_+ - 2\epsilon_-\ .
\end{align}
We indeed find that $\epsilon_+$ defines an $\mathcal{N}=1$ supersymmetry on this (free) QM model, while $\epsilon_-$ describes a trivial shift symmetry for its Goldstino mode $\chi_-$.

\subsection{$\eta\to 0$ in the quantum theory}

Our focus in analysing this $(1+1)$-dimensional model has been to develop a toy model for the conceptually comparable but technically much more difficult example of the non-Lorentzian 5d model for the M5-brane \cite{Lambert:2010wm,Lambert:2018lgt}, as discussed in Section \ref{sec: SYM}. In particular, in considering the $\eta\to 0$ limit we have localised exactly to the soliton moduli space, and thus have neglected possible quantum corrections to the resulting quantum mechanical model. While the focus of this paper is this classical construction, it is worth briefly discussing how one can include quantum corrections to this procedure, so as to make closer contact with the existing literature on the quantisation of fields around solitons \cite{Weinberg:2006rq}. In particular, we shall see that by taking the $\eta\to 0$ limit in the full quantum theory, we will generically have 1-loop corrections to the final QM action corresponding to loops of the fluctuations transverse to the moduli space. Further, we will see that 1-loop corrections also arise if we instead start with the fixed point action (\ref{eq: kink fixed point action}), and that (at least in this simple example) these corrections match those of the former calculation. \\

We first consider the partition function of the theory (\ref{eq: rescaled kink action}) at finite $\eta$, analytically continued to Euclidean signature,
\begin{align}
  \mathcal{Z}_\eta = \int D\phi\, D\psi\, \exp\Big\{-\left( \eta^{-1} S_{-1} + S_0 + \eta S_1 \right)\Big\}\ .
\end{align}
Then, as is familiar from localization calculations \cite{Cremonesi:2014dva}, we can determine the $\eta\to 0$ limit of $\mathcal{Z}_\eta$ by treating $\eta$ as a semiclassical expansion parameter - sometimes referred to as an auxiliary Planck constant\footnote{It is implicit that the overall quantum parameter $\hbar$ has been set to 1}. In other words, we expand in $\eta$ around the (moduli space of) saddle points of $S_{-1}$. 

We choose boundary conditions such that we lie in a particular kink sector. In other words, we fix $\phi$ at $x=\pm\infty$, and assume $W(\phi(\infty))-W(\phi(-\infty))>0$. Then, to describe the QFT around the classical minima of such a sector, we shift $S_{-1}$ as
\begin{align}
  S_{-1} \to S_{-1} - [W(\phi(\infty))-W(\phi(-\infty))] = \frac{1}{2} \int d^2 x\,\, g_{ij} \left( \partial_x\phi^i - D^i W \right) \left( \partial_x\phi^j - D^j W \right)\ ,
\end{align}
where the overall sign difference from (\ref{eq: rescaled kink action}) arises from the Wick rotation to Euclidean signature. Now let $\phi_\text{cl}^i(x; m^\alpha(t))$ be a solution for $\phi^i$ lying on the saddle point locus of $S_{-1}$, i.e. $\partial_x\phi_\text{cl}^i - D^i W(\phi_\text{cl})=0$. Here, the functions $m^\alpha(t)$ denote the (time-dependant) moduli. We then expand,
\begin{align}
  \phi^i(t,x) = \phi_\text{cl}^i(x; m^\alpha(t)) + \eta^{1/2}\delta \phi^i\ ,
\end{align}
where the perturbations $\delta \phi^i$ lie transverse to the moduli space,
\begin{align}
  \int dx\,\,g_{ij} \delta \phi^i \frac{\partial \phi^j_\text{cl}}{\partial m^\alpha}=0\ .
\end{align}
Then, we find
\begin{align}
  \mathcal{Z}_\eta = \int Dm\, D(\delta \phi)\, D\psi \exp \left\{ -S_0[\phi_\text{cl}] -\frac{1}{2} \int d^2 x\,\, \delta \phi^i \left( \frac{\delta^2 S_{-1}}{\delta\phi^i \delta\phi^j}[\phi_\text{cl}] \right) \delta \phi^j + \mathcal{O}(\eta^{1/2}) \right\}
\end{align}
Hence, taking $\eta\to 0$, we have
\begin{align}
  \mathcal{Z}_0 = \int Dm\, D\psi \,\mathcal{Z}_\text{1-loop}\, e^{-S_0[\phi_\text{cl}]}\ ,
\end{align}
where
\begin{align}
  \mathcal{Z}_\text{1-loop} = \int D(\delta \phi) \exp \left\{ -\frac{1}{2} \int d^2 x\,\, \delta \phi^i \left( \frac{\delta^2 S_{-1}}{\delta\phi^i \delta\phi^j}[\phi_\text{cl}] \right) \delta \phi^j \right\} = \left[ \det \left( \frac{\delta^2 S_{-1}}{\delta\phi^i \delta\phi^j}[\phi_\text{cl}] \right) \right]^{-1/2}\ .
\end{align}
Hence, the classical moduli space action $S_0[\phi_\text{cl}]$, precisely as discussed in section \ref{subsec: reduction to constraint surface} up to a Wick rotation, receives quantum corrections from $\mathcal{Z}_\text{1-loop}$, which we have neglected in our classical analysis. In particular, setting $g_{ij}=\delta_{ij}$ for simplicity, we have
\begin{align}
  \frac{\delta^2 S_{-1}}{\delta\phi^i \delta\phi^j} [\phi_\text{cl}] = -\delta_{ij}\partial_x^2 +( \partial_k \partial_i \partial_j W )(\phi_\text{cl})\,\partial_x\phi^k_\text{cl} + (\partial_i \partial_k W)(\partial_j \partial_k W)\ .
\end{align}
It is natural then to ask how we should interpret the QFT defined by what we called the `fixed point' action $\tilde{S}$ (\ref{eq: kink fixed point action}). We can once again consider the path integral, this time remaining in Lorentzian signature,
\begin{align}
  \tilde{\mathcal{Z}} = \int D\phi\, D\psi\, DG\, e^{i\tilde{S}[\phi,\psi]} = \int D\phi\, D\psi\, DG\,\exp\left\{ i S_0[\phi,\psi] + i\int d^2 x\, G_i \left( \partial_x\phi^i - D^i W \right)  \right\}\ .
\end{align}
Integrating out the Lagrange multiplier $G_i$, we arrive at
\begin{align}
  \tilde{\mathcal{Z}}_0 = \int D\phi\, D\psi\,\, \delta\left( \partial_x\phi^i - D^i W \right) e^{iS_0[\phi,\psi]} = \int Dm\, D\psi\, \tilde{\mathcal{Z}}_\text{1-loop} e^{iS_0[\phi_\text{cl},\psi]}\ ,
\end{align}
where, again taking $g_{ij}=\delta_{ij}$ for simplicity,
\begin{align}
  \tilde{\mathcal{Z}}_\text{1-loop} = \left[\det \Big( \delta_{ij} \partial_x - \partial_i\partial_j W (\phi_\text{cl})  \Big) \right]^{-1}\ .
\end{align}
Once again, we find that the action constrained exactly to the moduli space $S_0[\phi_\text{cl}]$ receives corrections in the form of a one-loop determinant. Indeed, it is not hard to see that we have $\mathcal{Z}_\text{1-loop}=\tilde{\mathcal{Z}}_\text{1-loop}$. Defining $\Delta_{ij}=\delta_{ij} \partial_x - \partial_i\partial_j W (\phi_\text{cl})$, we have that
\begin{align}
  \left( \Delta^\dagger \Delta \right)_{ij} &= \left( -\delta_{ik} \partial_x - \partial_i \partial_k W(\phi_\text{cl}) \right)\left(\delta_{kj} \partial_x - \partial_k \partial_j W(\phi_\text{cl}) \right)\nonumber\\
  &= -\delta_{ij}\partial_x^2 +( \partial_k \partial_i \partial_j W )(\phi_\text{cl})\,\partial_x\phi^k_\text{cl} + (\partial_i \partial_k W)(\partial_j \partial_k W) \nonumber\\
  &= \frac{\delta^2 S_{-1}}{\delta\phi^i \delta\phi^j} [\phi_\text{cl}]\ ,
\end{align}
and so
\begin{align}
  \mathcal{Z}_\text{1-loop} = \left[\det \left( \Delta^\dagger \Delta \right)\right]^{-1/2} = \left[\det ( \Delta^\dagger ) \det \left( \Delta \right) \right]^{-1/2} = \left[ \det \left( \Delta \right) \right]^{-1} = \tilde{\mathcal{Z}}_\text{1-loop}\ .
\end{align}
Hence we have shown, at least in this simple kink example and with flat target space, that the one-loop corrections to the moduli space action are independent of our choice to either consider the $\eta\to 0$ limit in the full quantum theory, or to just use the fixed point action $\tilde{S}$ to begin with.

%
%

\section{Instantons in 5d super-Yang-Mills, and the M5-brane}\label{sec: SYM}

We now turn our attention to a non-Lorentzian Yang-Mills-like theory in 5 dimensions, given by the action \cite{Lambert:2018lgt,Lambert:2019nti}
\begin{align}
  S = \frac{1}{g^2} \text{tr}\int d^4 x\, dx^0\, \Bigg( \frac{1}{2} F_{0i} F_{0i} + \frac{1}{2} F_{ij} G_{ij} - \frac{1}{2} \left( D_i X^I \right) \left( D_i X^I \right) \nonumber\\
  - \frac{i}{2} \bar{\Psi}_+ \Gamma_- D_0 \Psi_+  +i \bar{\Psi}_- \Gamma_i D_i \Psi_+ + \frac{1}{2} \bar{\Psi}_+ \Gamma_- \Gamma^I [X^I,\Psi_+] \Bigg)\ .
  \label{eq: 5d action}
\end{align}
Here we have indices $i,j,...=1,2,3,4$ and $I,J,...=6,7,8,9,10$. The $\Gamma$-matrices form a real $32\times 32 $ representation of 11-dimensional Clifford algebra with signature $(-,+,\dots,+)$. Furthermore, we define $\Gamma_\pm = \tfrac{1}{\sqrt{2}}\left( \Gamma_0 \pm \Gamma_5 \right)$, and all spinors are taken to be real.

We have gauge field $A_\mu = \left( A_0, A_i \right)$ with $\mu=0,1,2,3,4$, with field strength $F_{\mu\nu}= \partial_\mu A_\nu - \partial_\nu A_\mu - i [A_\mu, A_\nu]$. The matter fields transform in the adjoint of the gauge group $G=SU(N)$. They are the 5 scalars $X^I$, $I=6,\dots,10$ and a single spinor field $\Psi$ satisfying $\Gamma_{012345}\Psi=-\Psi$. The $\pm$ subscript denotes chirality under $\Gamma_{05}=\Gamma_{-+}$, i.e. $\Psi_\pm = \tfrac{1}{2}\left( 1\pm \Gamma_{05} \right) \Psi$. The spatial 2-form $G_{ij}$ is anti-self-dual; $G_{ij} = -\tfrac{1}{2} \varepsilon_{ijkl}G_{kl}$.\\

The theory enjoys 16 rigid supersymmetries and a further 8 superconformal symmetries, which are neatly contained in the field transformations
\begin{align}
  \delta X^I &= i \bar{\epsilon}_+ \Gamma^I \Psi_- + i\bar{\epsilon}_- \Gamma^I \Psi_+ \nonumber \\
  \delta A_0 &= i \bar{\epsilon}_- \Psi_+ - i \bar{\epsilon}_+ \Psi_- \nonumber \\
  \delta A_i &= -i\bar{\epsilon}_+ \Gamma_i \Gamma_- \Psi_+ \quad \nonumber \\
  \delta \Psi_+ &= F_{0i} \Gamma_i \epsilon_+  + \tfrac{1}{4} F_{ij} \Gamma_{ij} \Gamma_+ \epsilon_- + \left( D_i X^I \right) \Gamma_i \Gamma^I \epsilon_+  \nonumber \\
  \delta \Psi_- &= -F_{0i} \Gamma_i \epsilon_- + \left( D_0 X^I \right) \Gamma^I \Gamma_- \epsilon_+ + \left( D_i X^I \right) \Gamma_i \Gamma^I \epsilon_-  \nonumber \\
  &\hspace{8mm}+ \tfrac{i}{2} [X^I, X^J] \Gamma^{IJ} \Gamma_- \epsilon_+ + \tfrac{1}{4} G_{ij} \Gamma_{ij} \Gamma_- \epsilon_+ -4 X^I \Gamma^I \zeta_- \nonumber\\
  \delta G_{ij} &= \tfrac{i}{2}\bar{\epsilon}_-\Gamma_+ \Gamma_k \Gamma_{ij} D_k \Psi_- - i\bar{\epsilon}_+ \Gamma_{ij}  D_0 \Psi_- - \bar{\epsilon}_+ \Gamma_{ij}  \Gamma^I [X^I, \Psi_-] + 3i\bar\zeta_-\Gamma_{+}\Gamma_{ij}\Psi_-\ ,
  \label{eq: SYM unshifted SUSY}
\end{align}
where
\begin{align}
  \epsilon=\xi+ \left( x^0 \Gamma_+ +x^i \Gamma_i \right) \zeta_- \implies \begin{cases}
 	\epsilon_+ &= \xi_+ + x^0 \Gamma_+ \zeta_- \\
 	\epsilon_- &= \xi_- + x^i \Gamma_i \zeta_- 
 \end{cases}
\end{align}
and $\Gamma_{012345}\xi = \xi$, $\Gamma_{012345}\zeta_- = -\zeta_-$. Then, neglecting boundary terms at temporal infinity, we have
\begin{align}
  \delta S = \frac{1}{g^2}\text{tr}\int d^5 x\,\, \partial_i \Big[\,\, &\bareps_+ \left( i F_{0i} \Psi_- - i \left( D_i X^I \right)\Gamma^I \Psi_- \right.\nonumber\\
  &\qquad  \left. + i \left( D_0 X^I \right) \Gamma_- \Gamma^I \Gamma_i \Psi_+ - \tfrac{1}{2} \left[ X^I, X^J \right] \Gamma_- \Gamma^{IJ} \Gamma_i \Psi_+ \right) \nonumber\\[0.8em]
  +\,\,&\bareps_- \left( -i F_{0j} \Gamma_{ij} \Psi_+ - i \left( D_j X^I \right) \Gamma^I \Gamma_{ij} \Psi_+ + \tfrac{i}{4} F_{jk} \Gamma_+ \Gamma_i \Gamma_{jk} \Psi_- \right) \nonumber\\[0.8em]
  +\,\, & 4i X^I \bar{\zeta}_- \Gamma^I \Gamma_i \Psi_+ \Big]\ .
\end{align}
Thus, for suitable boundary conditions we have $\delta S=0$.\\

The on-shell conditions of this theory were first derived as a special solution to a set of equations defining a representation of the 6-dimensional $(2,0)$-supersymmetric tensor multiplet \cite{Lambert:2010wm}, in the context of which the theory described a null compactification of a stack of M5-branes. Further analysis \cite{Lambert:2011gb} showed that the dynamics reduced to geodesic motion on instanton moduli space, thus leading to the light-cone proposal for the $(2,0)$ theory \cite{Aharony:1997an,Aharony:1997th}. However, this construction provided only the bosonic sector of the resulting QM model, while also specialising to spherically symmetric, commuting zero modes for the fields $A_0,X^I$.

Later work \cite{Lambert:2018lgt} determined the action (\ref{eq: 5d action}) giving rise to these on-shell conditions. It was then shown that the theory could be found as a scaling limit of 5d super-Yang-Mills, corresponding to a stack of space-like compactified M5-branes in the limit that the compact direction became null \cite{Lambert:2019nti}. It was further noted that in this limit, super(conformal)-symmetry was enhanced from 16 to 24 supercharges, a process further explored holographically \cite{Lambert:2019jwi}.

Evident from its construction as a scaling limit, $S$ enjoys a Lifshitz scaling symmetry. We introduce the notation $[\Phi]=\left( a,b \right)$ for an object $\Phi$, where $a$ is the mass dimension, and $b$ the Lifshitz scaling dimension. Then,
\begin{align}
  \left[ x^0 \right] = - \left[ A_0 \right] & = \left( -1 , -1 \right) & \left[ \Psi_+ \right] & = \left( \tfrac{3}{2} , 1 \right) \nonumber\\
  \left[ x^i \right] = - \left[ A_i \right]  & = \left( -1 , -\tfrac{1}{2} \right) & \left[ \Psi_- \right] & = \left( \tfrac{3}{2} , \tfrac{3}{2} \right) \nonumber\\
  \left[ X^I \right] & = \left( 1 , 1 \right) & \left[ \xi_+ \right] & = \left( -\tfrac{1}{2} ,-\tfrac{1}{2} \right) \nonumber\\
  \left[ G_{ij} \right] & = \left( 2 , 2 \right) & \left[ \xi_- \right] & = \left( -\tfrac{1}{2} , 0 \right) \nonumber\\
  \left[ 1/g^2 \right]& = \left( 1 , 0 \right)  & \left[ \zeta_- \right] & = \left( \tfrac{1}{2} , \tfrac{1}{2} \right)\ .
  \label{eq: scaling weights}
\end{align}

\subsection{Reduction to constraint surface}

We now reduce the theory by integrating out non-dynamical fields. Letting $E^{(\Phi)}=g^2 \frac{\delta S}{\delta \Phi}$ denote the on-shell condition corresponding to a field $\Phi$, we find
\begin{align}
  E^{\left( G_{ij} \right)} &= \tfrac{1}{2} F_{ij}^- \nonumber\\
  E^{\left( \Psi_- \right)} &= i \Gamma_i D_i \Psi_+ \nonumber\\
  E^{\left( X^I \right)} &= D_i D_i X^I - \bar{\Psi}_+ \Gamma_- \Gamma^I \Psi_+ \nonumber\\
  E^{\left( A_0 \right)} &= D_i F_{0i} + \bar{\Psi}_+ \Gamma_- \Psi_+ \label{eq: SYM constraints}\\[1em]
  E^{\left( A_i \right)} &= - D_0 F_{0i} + D_j G_{ij} + i \left[ X^I, D_i X^I \right] - \bar{\Psi}_+ \Gamma_i \Psi_- - \bar{\Psi}_- \Gamma_i \Psi_+  \nonumber\\
  E^{\left( \Psi_+ \right)} &= -i\Gamma_- D_0 \Psi_+ + i \Gamma_i D_i \Psi_- + \Gamma_- \Gamma^I \left[ X^I, \Psi_+ \right]\ ,
\end{align}
where we have assumed suitable (i.e. Dirichlet or Neumann) boundary conditions for $X^I$ and $A_0$ at spatial infinity. Indeed, it is the time-dependent parameters describing such boundary conditions that determines the energy of these fields, and hence they will generically appear in the reduced quantum mechanics. 

The first four of the equations (\ref{eq: SYM constraints}), corresponding to the non-dynamical fields in $S$, define the constraint surface $\Sigma=\left\{E^{\left( G_{ij} \right)}, E^{\left( \Psi_- \right)}, E^{\left( X^I \right)}, E^{\left( A_0 \right)}=0\right\}$. The $G_{ij}$ equation of motion restricts $A_i$ to the moduli space $\mathcal{I}_{k,N}$ of a degree $k$ instanton for some $k\ge 0$. In particular, we have $\dim\left( \mathcal{I}_{k,N} \right)=4kN$ \cite{Belitsky:2000ws}. The $\Psi_-$ equations restricts $\Psi_+$ to solve the gauge covariant Dirac equation. Using index theorem techniques, we find that the moduli space of normalisable solutions is described by $8kN$ Grassmann parameters. This calculation follows the standard argument \cite{Belitsky:2000ws}, but crucially differs by a factor of 4 from the standard result of $2kN$ for an adjoint fermion in four dimensions due to our use of 32-component spinors of $\text{Spin}(1,10)$.\footnote{Usually, the relevant index is proportional to $\text{tr}(\gamma_5 \gamma_\mu\gamma_\nu\gamma_\rho\gamma_\sigma)=4\varepsilon_{\mu\nu\rho\sigma}$ where the $\{\gamma_\mu\}_{\mu=1}^4$ form a basis for the 4d Euclidean Clifford algebra, and the fermionic zero modes in the background of an instanton (rather than anti-instanton) have negative chirality under $\gamma_5=\gamma_1\gamma_2\gamma_3\gamma_4$. The equivalent object in our formulation is $\text{tr}(\Gamma_{1234}\Gamma_i\Gamma_j\Gamma_k\Gamma_l)=32\varepsilon_{ijkl}$. We indeed have $\Gamma_{1234}\Psi_+ = -\Psi_+$, and so have non-trivial zero modes. Taking into account the additional chirality condition on $\Psi_+$ under $\Gamma_{05}$, we find $\frac{1}{2}\times \frac{32}{4}=4$ times as many zero modes.}


The kinetic term for $\Psi_+$ in $S$ implies kinetic terms for these Grassmann parameters in the reduced theory. Hence, on-shell in the reduced theory, there are $8kN/2=4kN$ Grassmann degrees of freedom, and thus Bose-Fermi degeneracy is recovered on-shell. The remaining two equations restrict the non-dynamical $X^I$ and $A_0$ to the solution space of gauge covariant Poisson equations, and thus will be determined up a set of zero modes.\\

Our aim is now to constrain the theory to $\Sigma$ without breaking any supersymmetry. We find that under the supersymmetry (\ref{eq: SYM unshifted SUSY}), the constraints transform as
\begin{align}
  \delta E^{\left( G_{ij} \right)} &= -\tfrac{1}{4}\bareps_+ \Gamma_- \Gamma_{ij} E^{\left( \Psi_- \right)}  \nonumber\\
  \delta E^{\left( \Psi_- \right)} &= i \left( E^{\left( X^I \right)}\Gamma^I + E^{\left( A_0 \right)} \right)\epsilon_+  \nonumber\\
  \delta E^{\left( X^I \right)} &= \bareps_+ \left( \Gamma^I \Gamma_i D_i E^{\left( \Psi_+ \right)} + \Gamma_- \left( \Gamma^I D_0 - i \Gamma^{IJ}\left[ X^J, \,\cdot\,  \right] \right) E^{\left( \Psi_- \right)} - \Gamma^I \Gamma_{ij} \left[ E^{\left( G_{ij} \right)},\Psi_- \right] \right) \nonumber\\
   & \qquad + \bareps_- \Gamma^I \Gamma_i D_i E^{\left( \Psi_- \right)} + 2 \bar{\zeta}_- \Gamma^I E^{\left( \Psi_- \right)}  \nonumber\\
  \delta E^{\left( A_0 \right)} &= \bareps_+ \left( \Gamma_i D_i E^{\left( \Psi_+ \right)} + i \Gamma_- \Gamma^I \left[ X^I, E^{\left( \Psi_- \right)} \right] - \Gamma_{ij} \left[ E^{\left( G_{ij} \right)},\Psi_- \right] \right) \nonumber\\
   & \qquad - \bareps_-  \Gamma_i D_i E^{\left( \Psi_- \right)} + 3 \bar{\zeta}_- E^{\left( \Psi_- \right)}\ .
\end{align}
In particular, we note that $\left.\delta E^{(X^I)}\right|_\Sigma$ and $\left.\delta E^{(A_0)}\right|_\Sigma$ are generically non-zero for $\epsilon_+$ non-zero, and thus only the supersymmetry $\xi_-$ is unbroken by restricting to $\Sigma$.\\

We now show that this issue can be remedied such that the surface $\Sigma$ preserves the full 24 supercharges. We introduce a shifted supersymmetry $\hat{\delta}$, defined by
\begin{align}
  \hat\delta X^I &= \delta X^I -i\bar{\epsilon}_+ \Gamma^I \Psi_- - \bareps_+ \Gamma_- \Gamma^I \chi + \phi^I_{(X)} &=&\,\,  i\bar{\epsilon}_- \Gamma^I \Psi_+ - \bareps_+ \Gamma_- \Gamma^I \chi + \phi^I_{(X)} \nonumber \\
  \hat\delta A_0 &= \delta A_0 + i \bar{\epsilon}_+ \Psi_- - \bareps_+\Gamma_- \chi + \phi_{(A)} &=&\,\,  i \bar{\epsilon}_- \Psi_+ - \bareps_+\Gamma_- \chi + \phi_{(A)}\ ,
  \label{eq: shifted SUSY}
\end{align}
with $\hat{\delta}=\delta$ on other fields. Here, $\chi$ is a spinorial field satisfying
\begin{align}
  D_i D_i \chi = \Gamma_i \left[ F_{0i}, \Psi_+ \right] + \Gamma_i \Gamma^I \left[ D_i X^I, \Psi_+ \right]\ ,
  \label{eq: chi def}
\end{align}
while $\phi^I_{(X)},\phi_{(A)}$ are each solutions to the gauge-covariant Laplace equation, i.e. $D_i D_i \phi^I_{(X)}=0$, $ D_i D_i \phi_{(A)}=0$, which effectively allow for different boundary conditions on $\chi$. Then, it is easily seen that we have $\left.\hat{\delta} E^{(X^I)}\right|_\Sigma=0$, $\left.\hat{\delta} E^{(A_0)}\right|_\Sigma=0$ and thus we can restrict to $\Sigma$ without breaking any supersymmetry. We then calculate (again neglecting boundary terms at temporal infinity)
\begin{align}
  \hat{\delta} S = &\frac{1}{g^2} \text{tr}\int d^4 x\, dx^0\,\, \left( \left( i\bareps_+\Psi_- - \bareps_+\Gamma_-\chi + \phi_{(A)} \right) E^{\left( A_0 \right)} \right.\nonumber\\
  &\hspace{30mm}\left. - \left( i\bareps_+\Gamma^I\Psi_- + \bareps_+\Gamma_-\Gamma^I\chi - \phi^I_{(X)} \right)E^{\left( X^I \right)}  \right) \nonumber\\
  +&\frac{1}{g^2}\text{tr}\int d^4 x\, dx^0\,\, \partial_i \Big[\,\, \bareps_+ \left(  F_{0i} \Gamma_- \chi + \left( D_i X^I \right)\Gamma_-\Gamma^I \chi \right.\nonumber\\
  &  \hspace{34mm}\left. + i \left( D_0 X^I \right) \Gamma_- \Gamma^I \Gamma_i \Psi_+ - \tfrac{1}{2} \left[ X^I, X^J \right] \Gamma_- \Gamma^{IJ} \Gamma_i \Psi_+ \right) \nonumber\\[0.8em]
  &\hspace{21mm}+\bareps_- \left( -i F_{0j} \Gamma_{ij} \Psi_+ - i \left( D_j X^I \right) \Gamma^I \Gamma_{ij} \Psi_+ + \tfrac{i}{2} E^{(G_{jk})} \Gamma_+ \Gamma_i \Gamma_{jk} \Psi_- \right) \nonumber\\[0.8em]
  &\hspace{21mm}+  4i X^I \bar{\zeta}_- \Gamma^I \Gamma_i \Psi_+ - F_{0i} \phi_{(A)} - \left( D_i X^I \right) \phi^I_{(X)} \Big]\ .
  \label{eq: delta hat S}
\end{align}
Thus, we find that for suitable boundary conditions at spatial infinity, and choices of $\phi_{(X)}^I, \phi_{(A)}$, we have $\hat{\delta}S|_\Sigma =0$.\\

To proceed, we need to write down the general solution for $A_i$ in some gauge, as described by the ADHM construction \cite{Atiyah:1978ri}. One can then seek the general solutions to the three remaining constraints in terms of this ADHM data. The quantum mechanical action for the reduced theory $S_\Sigma$ is then determined by evaluating $S$ on these solutions and performing the spatial integral. In particular, the bosonic sector of the theory will reproduce the standard $\sigma$-model with the usual moduli space metric as given by
\begin{align}
  \mathcal{G}_{\alpha \beta} = \text{tr}\int d^4 x\,\,  \left( \delta_\alpha A_i  \right)\left( \delta_\beta A_i  \right) .
\end{align}
where $\alpha$ runs over the $4kN$ moduli of $A_i$, and we have fixed the time evolution of $A_i$ to lie transverse to gauge orbits. The full $S_\Sigma$ hence must be an extension to the maximal $\mathcal{N}=(4,4)$ $\Sigma$-model on instanton moduli space $\mathcal{I}_{k,N}$ to include coupling to the zero modes of $X^I$ and $A_0$. The theory will possess in total 24 supersymmetries, made up of the regular 8 rigid supersymmetries contained in $\xi_+$, their 8 superconformal partners in $\zeta_-$, and finally an additional 8 contained in $\xi_-$ which generically act only on the fermions and zero modes.

\subsection{The single $SU(2)$ instanton}

We now focus on the particular choice of gauge group $G=SU(2)$, and on the single instanton ($k=1$) sector. In the M-theory picture, this corresponds to a single unit of momentum along the compact null direction \cite{Aharony:1997an,Aharony:1997th}. This particular sector and gauge group are special, as they allow us to generate \textit{all} $8kN=16$ Grassmann moduli of $\Psi_+$ using an infinitesimal super(conformal) perturbation from a purely bosonic solution for the constraints (\ref{eq: SYM constraints}). For higher $k$ and/or $N$, one must seek the remaining $8(kN-2)$ zero modes via other means. 

\subsubsection{Solving the constraints}

We solve the constraints (\ref{eq: SYM constraints}) defining the constraint surface $\Sigma$, first finding a purely bosonic solution. The general solution for the instanton equation in the $k=1$ sector is
\begin{align}
  A_i(t,x) = h \hat{A_i} h^{-1} - i \left( \partial_i h \right) h^{-1}\ ,
\end{align}
where $\hat{A}_i$ is the $k=1$ instanton in singular gauge \cite{Belavin:1975fg},
\begin{align}
  \hat{A_i}(t,x) = \frac{\rho^2}{\left( x-y \right)^2 \left( \left( x-y \right)^2 + \rho^2 \right)} \bar{\eta}_{ij}^a \left( x-y \right)_j g\sigma^a g^{-1} \ .
\end{align}
Here, we have \textit{time-dependant} moduli $y_i(t)\in\mathbb{R}^4$, $\rho(t)\in\mathbb{R}$, and $g(t)\in SU(2)$, where we write $t=x^0$. Additionally, we have allowed our solution to move freely along gauge orbits over time, as signified by the gauge parameter $h:\mathbb{R}^{1,4}\to SU(2)$. We now choose the time evolution of $h$ such that the dynamical degrees of freedom of $A_i$ lie transverse to gauge orbits, and as such the resulting model describes only these physical degrees of freedom. This amounts to requiring that at any fixed time,
\begin{align}
  \text{tr}\int_{t=t_0}d^4 x\,\, \left( D_i\Lambda \right)\partial_0 A_i = 0\ ,
  \label{eq: transverse to gauge orbits}
\end{align}
for any $i\Lambda:\mathbb{R}^{1,4}\to \frak{su}(2)$ that vanishes at spatial infinity. This final stipulation ensures that $g(t)$ remains intact as a physical degree of freedom \cite{Tong:2005un}. Writing $i\dot{\omega} = h^{-1} \dot{h}\in\frak{su}(2)$, (\ref{eq: transverse to gauge orbits}) is equivalent to
\begin{align}
  \hat{D}_i\hat{D}_i\dot{\omega} = -\hat{D}_i \left( \partial_0 \hat{A}_i \right)\ ,
\end{align}
where $\hat{D}_i$ denotes the gauge covariant derivative with respect to $\hat{A}_i$. Introducing the notation $z_i(t,x) = x_i - y_i(t)$ to denote the spatial displacement from the instanton centre at time $t$, we have the solution
\begin{align}
  \dot{\omega} = \frac{\rho^2}{z^2 \left( z^2 + \rho^2 \right)} \left( \bar{\eta}_{ij}^a \dot{y}_i z_j - \dot{u}^a z^2 \right) g \sigma^a g^{-1}\ ,
\end{align}
where we define $\dot{u}^a$ by $g^{-1}\dot{g}=i\dot{u}^a \sigma^a$. This in turn determines $h$ for all time, given initial value $h_0= h(t_0)$ on some time slice $t=t_0$. Note, we could always shift $\dot{\omega}$ by some zero mode of $\hat{D}_i \hat{D}_i$, however, we shall see that such zero modes can be absorbed into the general solution for $A_0$.\\

We now turn to the other equations. Writing $X^I = h \hat{X}^I h^{-1}$, we find
\begin{align}
  E^{\left( X^I \right)}=0 \implies \hat{D}_i \hat{D}_i \hat{X}^I = 0 \ ,
\end{align}
while, noting that $D_i F_{0i} = D_i\left( \partial_0 A_i - D_i A_0 \right) = -D_i D_i A_0$, and writing $A_0 = h\hat{A}_0 h^{-1}$, we have
\begin{align}
  E^{\left( A_0 \right)}=0 \implies \hat{D}_i \hat{D}_i \hat{A}_0 = 0\ .
\end{align}
Hence, we can write our bosonic solution as
\begin{align}
  A_i &= h\hat{A}_i h^{-1} - i \left( \partial_i h \right)h^{-1} \nonumber \\
  A_0 &= h \hat{A}_0 h^{-1} \nonumber\\
  X^I &= h \hat{X}^I h^{-1}\ ,
\end{align}
where $\hat{X}^I$ and $\hat{A}_0$ are zero modes of the gauge covariant Laplacian in the background of the $k=1$ instanton $\hat{A}_i$ in singular gauge. By performing a further \textit{time-independent} gauge transformation, it is easily seen that $h_0$ can be arbitrarily fixed by gauge transformations, and thus will not appear in any gauge invariant. Finally, we perform a gauge transformation to bring our bosonic solution to a more convenient form,
\begin{align}
  A_i &= \hat{A}_i \nonumber\\
  A_0 &= \hat{A}_0 - \dot{\omega} \nonumber\\
  X^I &= \hat{X}^I\ ,
  \label{eq: bosonic solution}
\end{align}
with
\begin{align}
  \hat{A}_i &= \frac{\rho^2}{z^2 \left( z^2 + \rho^2 \right)} \bar{\eta}_{ij}^a z_j g\sigma^a g^{-1} \nonumber\\
  h^{-1} \dot{h} &= i \dot{\omega} =  i\frac{\rho^2}{z^2 \left( z^2 + \rho^2 \right)} \left( \bar{\eta}_{ij}^a \dot{y}_i z_j - \dot{u}^a z^2 \right) g \sigma^a g^{-1}\ .
\end{align}
We indeed see that any zero modes we could have added to $\dot{\omega}$ could be absorbed into $\hat{A}_0$.\\

With a bosonic solution in hand, we seek a general solution to the constraints (\ref{eq: SYM constraints}). In principle, starting from our bosonic solution, we can construct a solution with fermions turned on by performing a \textit{finite} supersymmetry transformation - sometimes referred to as the sweeping procedure \cite{Belitsky:2000ws}. Then, each field $\Phi$ is given by

\begin{align}
  \Phi = \left.\left( e^{\hat{\delta}}\Phi \right)\right|_{\text{bos. solution}} = \sum_{n=0}^\infty \frac{1}{n!} \left.\left( \hat{\delta}^n \Phi \right)\right|_{\text{bos. solution}}\ .
\end{align}
What's more, since $\hat{\delta}$ is parameterised by a finite number of Grassmann parameters, this series will necessarily terminate. However, this approach is cumbersome, and requires us to solve (\ref{eq: chi def}) for $\chi$ for generic configurations, and then calculate $\hat{\delta}\chi, \hat{\delta}^2\chi, \dots$.

Thankfully, there is a more straightforward approach we can take. First, we fix $A_i = \hat{A}_i$. Then, we note that to generate a non-trivial solution to $E^{\left( \Psi_- \right)}=i\Gamma_i D_i \Psi_+=0$ we need only consider an \textit{infinitesimal} variation of the bosonic solution (\ref{eq: bosonic solution}). To see this, note that on $\Sigma$ we have $\hat{\delta} E^{\left( \Psi_- \right)}=0$. In particular,
\begin{align}
  0 &= \left.\left( \hat{\delta} E^{\left( \Psi_- \right)} \right)\right|_{\text{bos. solution}} \nonumber\\
  &= \left.\left( i\Gamma_i D_i \deltahat \Psi_+ + \Gamma_i \left[ \deltahat A_i , \Psi_+ \right] \right)\right|_{\text{bos. solution}} \nonumber\\
  &= i \Gamma_i D_i \left.\left( \deltahat \Psi_+ \right)\right|_{\text{bos. solution}}\ .
\end{align}
And thus, $\deltahat \Psi_+$ evaluated on the bosonic solution provides a solution for the $E^{\left( \Psi_- \right)}=0$, for gauge field $A_i$ unchanged. Performing this calculation, we find the solution
\begin{align}
  \Psi_+ = F_{ij} \Gamma_{ij} \left( \alpha + z_k \Gamma_k \beta \right) = F_{ij} \Gamma_{ij} \alpha + 4 F_{ij} z_j \Gamma_i \beta\ ,
\end{align}
where $\alpha,\beta$ are Majorana spinors of $\text{Spin}(1,10)$ with $\Gamma_{05}\alpha = -\Gamma_{012345}\alpha = \alpha$, $\Gamma_{05}\beta = \Gamma_{012345}\beta = \beta$. In particular, $\alpha$ and $\beta$ together make up $8kN=16$ Grassmann parameters, and thus we have the general solution for $\Psi_+$.\\

Solutions for the final two constraints for $X^I$ and $A_0$ cannot be generated in the same way, as they are both quadratic in $\Psi_+$. Instead, we solve them by inspection. It is convenient to write the solutions as
\begin{align}
  X^I &= \hat{X}^I + \tilde{X}^I \nonumber\\
  A_0 &= \hat{A}_0 - \dot{\omega} + \tilde{A}_0\ ,
\end{align}
where as before $D_i D_i \hat{X}^I=0$, $D_i D_i \hat{A}_0=0$, and
\begin{align}
  \tilde{X}^I &= -\frac{\left( z^2 + \rho^2 \right)^2}{16\rho^2} \bar{\Psi}_+ \Gamma_-\Gamma^I \Psi_+ = i F_{ij} \left( \bar{\alpha} \Gamma_- \Gamma^I \Gamma_{ij} \alpha + 8 z_j \bar{\alpha} \Gamma_- \Gamma^I \Gamma_i \beta + 4z_j z_k \bar{\beta} \Gamma_- \Gamma^I \Gamma_{ik} \beta \right) \nonumber\\
  \tilde{A}_0 &= -\frac{\left( z^2 + \rho^2 \right)^2}{16\rho^2} \bar{\Psi}_+ \Gamma_- \Psi_+ = i F_{ij} \left( \bar{\alpha} \Gamma_- \Gamma_{ij} \alpha + 8 z_j \bar{\alpha} \Gamma_-  \Gamma_i \beta - 4z_j z_k \bar{\beta} \Gamma_- \Gamma_{ik} \beta \right)\ .
  \label{eq: tilde defs}
\end{align}\\[2em]
In summary, we find the general solution to (\ref{eq: SYM constraints}) given by
\begin{align}
  A_i &= \frac{\rho^2}{z^2 (z^2 + \rho^2)} \etabar^a_{ij} z_j\,\, g \sigma^a g^{-1} \nonumber\\
  \Psi_+ &= F_{ij} \Gamma_{ij} \left( \alpha + z_k \Gamma_k \beta \right) \nonumber\\
  X^I &= \hat{X}^I - \frac{(z^2+\rho^2)^2}{16\rho^2}\bar{\Psi}_+\Gamma_-\Gamma^I \Psi_+ \nonumber\\
  A_0 &= \hat{A}_0 + \frac{\rho^2}{z^2(z^2+\rho^2)}\left( z^2 \dot{u}^a - \etabar^a_{ij} \dot{y}_i z_j \right) g \sigma^a g^{-1} - \frac{(z^2+\rho^2)^2}{16\rho^2}\bar{\Psi}_+\Gamma_- \Psi_+\ .
  \label{eq: general solution}
\end{align}\\
Finally, we need to solve for $\chi$ in order to determine the supersymmetry variations of the moduli parameterising the constraint space $\Sigma$. We can now split the terms in (\ref{eq: chi def}) into parts linear and cubic in $\Psi_+$, as
\begin{align}
  D_i D_i \chi = \Gamma_i\, [ \partial_0 A_i - D_i \hat{A}_0 + D_i \dot{\omega}, \Psi_+ ] + \Gamma_i \Gamma^I [ D_i \hat{X}^I, \Psi_+] - \Gamma_i [ D_i \tilde{A}_0, \Psi_+] + \Gamma_i \Gamma^I [D_i \tilde{X}^I, \Psi_+ ]\ ,
\end{align}
with $\tilde{X}^I$ and $\tilde{A}_0$ the fermion bilinears as defined in (\ref{eq: tilde defs}). We find the solution
\begin{align}
  \chi = &\,\,\hat{\chi} -\frac{\left( z^2+\rho^2 \right)^2}{16\rho^2}\Gamma_i\, [ \partial_0 A_i + D_i \dot{\omega}, \Psi_+ ] \nonumber\\
  &+\frac{8\rho^2}{z^2\left( z^2 + \rho^2 \right)^3}\bigg( 4 \etabar^a_{il} z_l z_j z_k \Big( \left( \bar{\alpha}\Gamma_- \Gamma_{ij} \alpha \right) \Gamma_k \alpha + \left( \bar{\alpha}\Gamma_-\Gamma^I \Gamma_{ij} \alpha \right) \Gamma^I\Gamma_k \alpha \Big) \nonumber\\
  &\hspace{32mm}+ 4\left( z^2 -\rho^2 \right)\etabar^a_{ik} z_k z_j \Big( \left( \bar{\alpha}\Gamma_- \Gamma_{ij} \alpha \right) \beta + \left( \bar{\alpha}\Gamma_-\Gamma^I \Gamma_{ij} \alpha \right) \Gamma^I\beta \Big) \nonumber\\
  &\hspace{32mm}-4 z^2 \etabar^a_{ik}z_j z_l \Big( \left( \bar{\alpha}\Gamma_- \Gamma_{ij} \alpha \right) \Gamma_{kl}\beta + \left( \bar{\alpha}\Gamma_-\Gamma^I \Gamma_{ij} \alpha \right) \Gamma^I\Gamma_{kl}\beta \Big) \nonumber\\
  &\hspace{32mm}+ 4 \etabar^a_{il} z_l z^2 \left( \rho^2 \delta_{jk} - z_j z_k \right) \Big( \left( \bar{\beta}\Gamma_- \Gamma_{ij} \beta \right) \Gamma_{k}\alpha - \left( \bar{\beta}\Gamma_-\Gamma^I \Gamma_{ij} \beta \right) \Gamma^I\Gamma_{k}\alpha \Big) \nonumber\\
  &\hspace{32mm}+ \rho^2 z^4 \etabar^a_{ij} \Big( \left( \bar{\beta}\Gamma_- \Gamma_{ij} \beta \right) \beta - \left( \bar{\beta}\Gamma_-\Gamma^I \Gamma_{ij} \beta \right) \Gamma^I\beta \Big) \bigg)\,\, g \sigma^a g^{-1}\ ,
  \label{eq: chi solution}
\end{align}
where $\hat{\chi}$ is a solution to
\begin{align}
  D_i D_i \hat{\chi} = -\Gamma_i\, [ D_i \hat{A}_0 , \Psi_+ ] + \Gamma_i \Gamma^I [ D_i \hat{X}^I, \Psi_+]\ ,
  \label{eq: chi hat def}
\end{align}
for the so-far unspecified zero modes $\hat{X}^I$ and $\hat{A}_0$.

\subsubsection{The reduced theory}

Let $m^\alpha$ formally denote the set of moduli $y_i,\rho,u^a,\alpha,\beta$ combined with the time-dependant parameters describing the zero modes $\hat{X}^I$ and $\hat{A}_0$. We now want to determine the quantum mechanical action for the reduced theory, along with how supersymmetry acts on the $m^\alpha$. The action $S_\Sigma$ is determined by simply evaluating our original action $S$ on the solutions (\ref{eq: general solution}).

To determine how supersymmetry acts on the $m^\alpha$, we first need to augment the supersymmetry variation $\hat{\delta}$ by an infinitesimal gauge transformation with parameter $\tau$, such that the new variation lies transverse to gauge orbits. Indeed, we already did essentially the same thing in defining $\dot{\omega}$ to force time evolution to lie transverse to gauge orbits, and as such $\tau$ takes the same form. Then, letting $\delta_\Sigma$ denote the supersymmetry of the reduced theory, we have
\begin{align}
 \delta_\Sigma m^\alpha \left( \frac{\partial A_i}{\partial m^\alpha} \right) &= -i\bar{\epsilon}_+ \Gamma_i \Gamma_- \Psi_+ + D_i\tau  \nonumber \\
 \delta_\Sigma  m^\alpha \left( \frac{\partial \Psi_+}{\partial m^\alpha} \right) &= F_{0i} \Gamma_i \epsilon_+  + \tfrac{1}{4} F_{ij} \Gamma_{ij} \Gamma_+ \epsilon_- + \left( D_i X^I \right) \Gamma_i \Gamma^I \epsilon_+ + i [\tau,\Psi_+] \nonumber\\
  \delta_\Sigma  m^\alpha \left( \frac{\partial X^I}{\partial m^\alpha} \right) &= i\bar{\epsilon}_- \Gamma^I \Psi_+ - \bareps_+ \Gamma_- \Gamma^I \chi + \phi^I_{(X)} + i[\tau,X^I] \nonumber\\
    \delta_\Sigma  m^\alpha \left( \frac{\partial A_0}{\partial m^\alpha} \right) &= i \bar{\epsilon}_- \Psi_+ - \bareps_+\Gamma_- \chi + \phi_{(A)} +D_0 \tau \ ,
    \label{eq: reduced SUSY}
\end{align}
where
\begin{align}
  \tau = \hat{\tau}-\frac{\rho^2}{z^2 \left( z^2 + \rho^2 \right)}\Big( \etabar^a_{ij} \left( \delta_\Sigma y_i \right) z_j -z^2 \left( \delta_\Sigma  u^a \right) \Big)\,\, g \sigma^a g^{-1}\ ,
\end{align}
and $\hat{\tau}$ is any solution to $D_i D_i \hat{\tau}=0$, with different choices of $\hat{\tau}$ giving rise to different transformations of the $m^\alpha$ under $\delta_\Sigma$. Choosing $\hat{\tau}=0$, first equation of (\ref{eq: reduced SUSY}) is solved to find
\begin{align}
  \delta_\Sigma y_i &= 4i \left( \bareps_+ \Gamma_- \Gamma_i \alpha \right) \nonumber\\
  \delta_\Sigma \rho &= 4i\rho \left( \bareps_+ \Gamma_- \beta \right) \nonumber\\
  \delta_\Sigma u^a &= -i \etabar^a_{ij} \left( \bareps_+ \Gamma_- \Gamma_{ij} \beta \right)\ .
\end{align}
In particular, we note that the 16 supercharges contained in $\epsilon_+$ descend to full super(conformal) symmetries of the reduce quantum mechanical model, while the final 8 supercharges in $\xi_-$ will appear as fermionic shift symmetries.

To proceed, we need to write down the general \textit{finite energy} configurations for the zero modes $\hat{X}^I, \hat{A}_0$. One can then solve (\ref{eq: chi hat def}) for $\hat{\chi}$, and then seek solutions for $\phi_{(X)}^I$ and $\phi_{(A)}$ such that $\hat{\delta} S$ as calculated in (\ref{eq: delta hat S}) vanishes for our solutions (\ref{eq: general solution}). Finally, $\delta_\Sigma m^\alpha$ for the remaining moduli can be determined via (\ref{eq: reduced SUSY}), which will satisfy $\delta_\Sigma S_\Sigma=0$. For simplicity, we present a consistent truncation of the set of moduli $\{m^\alpha\}$ to include only the spherically symmetric and regular modes of $\hat{X}^I$, $\hat{A}_0$; this amounts to having chosen spherically symmetric (but time-dependant) Dirichlet boundary conditions for $X^I$, $A_0$. We note, however, that due to the opposite sign gradient terms for $X^I$ and $A_0$ in $S$, we generically expect additional higher modes whose energies cancel, but which may nonetheless couple to fermions in the reduced theory. Leaving such analysis to future work, we write the spherically symmetric solutions as
\begin{align}
  \hat{X}^I = \frac{z^2}{z^2+\rho^2} v^{I,a} g \sigma^a g^{-1},\qquad \hat{A}_0 = \frac{z^2}{z^2+\rho^2} w^a g \sigma^a g^{-1}\ ,
  \label{eq: sph symm modes}
\end{align}
for some $v^{I,a}(t), w^a(t)$, as first considered in \cite{Lambert:2011gb}. It is easily checked that these solutions do indeed contribute finite energy for general $v^{I,a}(t), w^a(t)$. These choices admit the solution
\begin{align}
  \hat{\chi}= -\frac{\left( z^2+\rho^2 \right)^2}{16\rho^2} \left( -\Gamma_i\, [ D_i \hat{A}_0 , \Psi_+ ] + \Gamma_i \Gamma^I [ D_i \hat{X}^I, \Psi_+] \right)\ .
\end{align}
for (\ref{eq: chi hat def}). Then, for these configurations (\ref{eq: sph symm modes}), we find
\begin{align}
  S_\Sigma = \frac{2\pi^2}{g^2} \int dt\,\, \bigg(& \dot{y}_i \dot{y}_i + 2 \dot{\rho}^2 + 2\rho^2 (\dot{u}^a - w^a)(\dot{u}^a - w^a) - 2\rho^2 v^{I,a} v^{I,a} \nonumber\\
  &-16 i \bar{\alpha} \Gamma_- \dot{\alpha} - 32\rho^2 i \bar{\beta} \Gamma_- \dot{\beta} \nonumber\\
  &- 8 \rho^2 i (\dot{u}^a - w^a)\eta^a_{ij} \bar{\beta} \Gamma_- \Gamma_{ij} \beta + 8 \rho^2 i v^{I,a} \eta^a_{ij} \bar{\beta} \Gamma_-\Gamma^I \Gamma_{ij} \beta \bigg)\ ,
  \label{eq: QM action}
\end{align}
while we find supersymmetries $\delta_\Sigma$ given by
\begin{align}
    \delta_\Sigma  y_i &= 4i \left( \bareps_+ \Gamma_- \Gamma_i \alpha \right) \nonumber\\[0.3em]
  \delta_\Sigma  \rho &= 4i\rho \left( \bareps_+ \Gamma_- \beta \right) \nonumber\\[0.3em]
  \delta_\Sigma  u^a &= -i \etabar^a_{ij} \left( \bareps_+ \Gamma_- \Gamma_{ij} \beta \right) \nonumber\\[0.3em]
  \delta_\Sigma \alpha &= \frac{1}{4} \dot{y}_i \Gamma_i \epsilon_+ - \frac{1}{4} y_i \Gamma_i \Gamma_+ \zeta_- + \frac{1}{4} \Gamma_+ \xi_- \nonumber\\[0.3em]
  \delta_\Sigma \beta &= \frac{\dot{\rho}}{4\rho}\epsilon_+ + \frac{1}{16} \etabar^a_{ij} (\dot{u}^a - w^a) \Gamma_{ij} \epsilon_+ + \frac{1}{16} \etabar^a_{ij} v^{I,a} \Gamma^I \Gamma_{ij} \epsilon_+ - \frac{1}{4}\Gamma_+ \zeta_- \nonumber\\[0.3em]
  & \qquad + 8i \left( \bareps_+ \Gamma_- \beta \right) \beta - i \left( \bareps_+ \Gamma_- \Gamma_{ij} \beta \right) \Gamma_{ij}\beta \nonumber\\[1em]
  \delta_\Sigma  v^{I,a} &= i\etabar^a_{ij} \left( \bareps_+ \Gamma_- \Gamma^I \Gamma_{ij} \dot{\beta} + \frac{2\dot{\rho}}{\rho} \bareps_+ \Gamma_- \Gamma^I \Gamma_{ij} \beta -2\bar{\zeta} \Gamma^I \Gamma_{ij} \beta  \right)\nonumber\\[0.3em]
  &\qquad -i \varepsilon^{abc}\etabar^b_{ij} v^{J,c} \EpsIJijBeta{I}{J}{i}{j} + i \varepsilon^{abc} \etabar^b_{ij} \left( \dot{u}^c - w^c \right) \EpsIijBeta{I}{i}{j} \nonumber\\
  &\qquad + 8\etabar^a_{ij} \Big( 2 \BetaijBeta{i}{j}\EpsIBeta{I} - \BetaijBeta{i}{k}\EpsIijBeta{I}{j}{k} \Big) \nonumber
\end{align}
\begin{align}
    \delta_\Sigma  w^a &= -i\varepsilon^{abc} \etabar^b_{ij} v^{I,c} \EpsIijBeta{I}{i}{j} \nonumber\\
    &\qquad + 8\etabar^a_{ij} \Big( 2\BetaijBeta{i}{k}\EpsijBeta{j}{k} - 2\BetaIijBeta{I}{i}{j}\EpsIBeta{I} \nonumber\\
    &\qquad\qquad\qquad   - \BetaIijBeta{I}{i}{k} \EpsIijBeta{I}{j}{k} \Big)\ ,
\end{align}
with
\begin{align}
  \epsilon_+ = \xi_+ + t \Gamma_+ \zeta_-\ .
\end{align}
Note, in verifying $\delta_\Sigma S_\Sigma=0$, it is helpful to use the Fierz relations
\begin{align}
  \BetaijBeta{i}{j}\BetaijBeta{i}{j}-\BetaIijBeta{I}{i}{j}\BetaIijBeta{I}{i}{j}&=0 \nonumber\\
  \BetaijBeta{i}{j}\BetaijBeta{i}{k}\EpsijBeta{j}{k}&=0
\end{align}
We recognise $S_\Sigma$ as an extension of the standard $\mathcal{N}=(4,4)$ $\sigma$-model, with a flat metric on the target space $\mathbb{R}^4\times \mathbb{R}^4$ (here, $u^a$ are the left-invariant $SU(2)$ forms of the unit $S^3$, which combined with radius $\rho$ gives a chart on $\mathbb{R}^4$). We note, however, that $g$ is indistinguishable from $-g$, and thus the actual target space is found by identifying $g\cong -g$ to find $\mathbb{R}^4 \times \left( \mathbb{R}^4/\mathbb{Z}_2 \right)$ which is indeed the hyper-K\"{a}hler moduli space of a single $SU(2)$ instanton. The model is extended by six $\frak{su}(2)$-valued time-dependant parameters $v^{I,a}$, $w^a$. We see that the full 8 rigid supersymmetries $\xi_+$ and 8 superconformal symmetries $\zeta_-$ now pair the bosonic and fermionic coordinates on $\Sigma$, while the rigid supersymmetry $\xi_-$ lives on as a shift symmetry for its Goldstino mode $\alpha$.\\

Naturally, the Lifshitz scaling symmetry of the initial theory (\ref{eq: 5d action}) descends to the quantum mechanics, where it is more naturally viewed simply as a conformal symmetry. In particular, in the notation of (\ref{eq: scaling weights}), we have
\begin{align}
  \left[ t \right] & = \left( -1 , -1 \right) & \left[ \alpha \right] & = \left( -\tfrac{1}{2} , 0 \right) \nonumber\\
  \left[ y_i \right]  & = \left( -1 , -\tfrac{1}{2} \right) & \left[ \beta \right] & = \left( \tfrac{1}{2} , \tfrac{1}{2} \right) \nonumber\\
  \left[ \rho \right] & = \left( -1 , -\tfrac{1}{2} \right) & \left[ \xi_+ \right] & = \left( -\tfrac{1}{2} ,-\tfrac{1}{2} \right) \nonumber\\
  \left[ u^a \right] & = \left( 0 , 0 \right) & \left[ \xi_- \right] & = \left( -\tfrac{1}{2} , 0 \right) \nonumber\\
  \left[ v^{I,a} \right]& = \left( 1 , 1 \right)  & \left[ \zeta_- \right] & = \left( \tfrac{1}{2} , \tfrac{1}{2} \right)\ . \nonumber\\
  \left[ w^a \right]& = \left( 1 , 1 \right) \nonumber\\
    \left[ 1/g^2 \right]& = \left( 1 , 0 \right)  
  \label{eq: QM scaling weights}
\end{align}\\
We also note that there is an additional class of symmetries acting on the parameters $v^{I,a},w^a$ corresponding to different solutions for $\phi^I_{(X)}$ and $\phi_{(A)}$. These symmetries are trivially local, as $v^{I,a},w^a$ appear only algebraically in $S$. For instance, we have $\delta_\sigma S=0$ for
\begin{align}
  \delta_a v^{I,a} &= \left( \bar{\sigma}_+\Gamma_- \Gamma^{I} \Gamma_{ij} \beta \right)\Big( i\varepsilon^{abc} \etabar^b_{ij} \left( \dot{u}^c - w^c \right) - 8 \etabar^a_{ik} \BetaijBeta{j}{k} \Big)\nonumber\\
    \delta_a w^a &= \left( \bar{\sigma}_+\Gamma_- \Gamma^{I} \Gamma_{ij} \beta \right)\Big( i\varepsilon^{abc} \etabar^b_{ij} v^{I,c} - 8 \etabar^a_{ik} \BetaIijBeta{I}{j}{k} \Big)\ ,
\end{align}
for spinor $\sigma_+(t)$ with $\Gamma_{012345}\sigma_+=\Gamma_{05}\sigma_+=\sigma_+$. Such symmetries can be better understood by considering the transformations
\begin{align}
  \delta_\lambda v^{I,a} &= \frac{v^{I,a} + 2i\etabar^a_{ij} \left( \bar{\beta} \Gamma_- \Gamma^I \Gamma_{ij} \beta \right)}{v^{J,b}v^{J,b} + 16 \left( \bar{\beta} \Gamma_-\Gamma_{kl} \beta \right)\left( \bar{\beta} \Gamma_- \Gamma_{kl} \beta \right)} \lambda_v(t) \nonumber\\
  \delta_\lambda w^a &= \frac{\left( \dot{u}^a - w^a \right) + 2i\etabar^a_{ij} \left( \bar{\beta} \Gamma_- \Gamma_{ij} \beta \right)}{\left( \dot{u}^b - w^b \right)\left( \dot{u}^b - w^b \right) + 16 \left( \bar{\beta} \Gamma_- \Gamma_{kl} \beta \right)\left( \bar{\beta} \Gamma_-  \Gamma_{kl} \beta \right)} \lambda_w(t)\ ,
\end{align}
where these expressions are understood as the corresponding (necessarily terminating) power series in $\beta$. Then, we have
\begin{align}
  \delta_\lambda S = \frac{2\pi^2}{g^2}\int dt\,\, \Big[ -4\rho^2 \left( \lambda_v + \lambda_w \right)  \Big]\ ,
\end{align}
and so in particular, $\delta_\lambda S=0$ for any $\lambda_w=-\lambda_v$. 

\section{Conclusion}\label{sec: conclusion}

In this paper, we presented a general procedure by which the dynamics of a particular class of non-Lorentzian supersymmetric field theories can be reduced to a supersymmetric quantum mechanical model. Such non-Lorentzian theories can be found as fixed points of induced RG flows \cite{Lambert:2019nti}, as well as from M-theory brane configurations both directly and holographically \cite{Lambert:2019jwi}. They generically involve a Lagrange multiplier imposing the Bogomol'nyi equation of a soliton preserving some amount of supersymmetry, and hence in the examples we have considered the resulting models are supersymmetric $\sigma$-models on the moduli space of these solitons.

We first considered the particularly simple and instructive example of a $\frac{1}{2}$-BPS kink in the $\mathcal{N}=(1,1)$ $\sigma$-model in $(1+1)$-dimensions with general potential, where we first considered a scaling limit and corresponding fixed point non-Lorentzian theory. Reducing to the constraint surface $\Sigma$, we determined a recipe for writing down the relevant supersymmetric $\sigma$-model on kink moduli space. We argued that this procedure provides a supersymmetric extension  to the standard scaling argument for the geodesic approximation for slow-moving solitons. Further, we discussed how one may calculate quantum corrections to this procedure.

We then revisited the non-Lorentzian $(4+1)$-dimensional Yang-Mills-like theory, which can be interpreted as a lightcone action for a stack of M5-branes. We demonstrated how dynamics can be reduced to motion on a constraint surface $\Sigma$ derived from integrating out non-dynamical fields. With some care taken to deform the supersymmetry algebra, we showed that the theory could be reduced to a supersymmetric $\sigma$-model on instanton moduli space, coupled to the zero modes of the embedding coordinates $X^I$ and electric potential $A_0$. We derived this $\sigma$-model explicitly for the case of the single $SU(2)$ instanton coupled only to spherically symmetric zero modes.\\

It would be interesting to consider non-Lorentzian RG flows and the subsequent soliton $\sigma$-models for other Yang-Mills solitons such as vortices and monopoles, especially when the parent theory includes coupling to matter. Here, one would expect to recover known results on the effective actions for these BPS solitons \cite{Gauntlett:1992yj,Gauntlett:1993sh,Cederwall:1995bc}, generically augmented by a set of couplings corresponding to matter fields of the parent theory.

It would also be interesting to generalise our analysis of the reduced theory in the SYM case, firstly to include all zero modes for $X^I$ and $A_0$, and then to use the ADHM formalism to investigate higher instanton number. Such analysis would in particular determine how the various modes of $X^I$ and $A_0$ appear in the reduced theory, and could thus shed light on how different M5-brane configurations are recovered in the $\sigma$-model. In doing such analysis, one may hope to recover results analogous to the D4-brane supertube \cite{Kim:2003gj} in the context of M5-branes. 

The quantum mechanical model (\ref{eq: QM action}), which constitutes an extension to the standard $\mathcal{N}=(4,4)$ $\sigma$-model with hyper-K\"{a}hler target, bears a resemblance to models obtained by considering the low energy limit of $\mathcal{N}=4$ SYM in four dimensions, where one of the six scalar vevs is taken to be much larger than the rest\cite{Weinberg:2006rq}. In particular, such a model also possesses an adjoint valued $SO(5)$ vector, akin to our $v^I$. It would be interesting to make closer contact with this construction, especially as it may help in constraining the form of the aforementioned generalisations.

Recent work \cite{Lambert:2019jwi} showed that the theory (\ref{eq: 5d action}) is in fact a special case of a non-Lorentzian M5-brane theory that can be derived holographically. In particular, one considers $AdS_7$ as a timelike fibration over a non-compact complex projective space $\tilde{\mathbb{CP}}_3$, places a stack of M5-branes at fixed $\tilde{\mathbb{CP}}_3$ radius, and then takes them to the boundary. The resulting theory is deformed from (\ref{eq: 5d action}) by a parameter $\Omega$, which in particular deforms the constraint imposed by $G_{ij}$, and crucially introduces a kinetic term for the $X^I$. It would be interesting to revisit this paper's construction for this theory, and understand the role of the $\Omega$ deformation in the resulting quantum mechanics. Further, one finds that the $\Omega$-deformed theory, and so therefore (\ref{eq: 5d action}), possesses a large set of bosonic symmetries corresponding to a subgroup of the conformal isometries of $AdS_7$ \cite{Lambert:2019fne}. One would expect these symmetries to descend to the quantum mechanical model.

Finally, it would be of interest to revisit the non-Lorentzian theories descended from the $(1+2)$-dimensional BLG and ABJM/ABJ Chern-Simons-matter theories, which describe a configuration of M2-branes that is U-dual to the lightcone M5-brane configuration discussed in this paper \cite{Kucharski:2017jwv,Lambert:2019nti}. The reduced theory would then be a $\sigma$-model on Hitchin moduli space \cite{Hitchin:1986vp}, again extended to include couplings to additional fields in the parent theory. We hope to report on these issues in due course.

\subsection*{Acknowledgements}

Many thanks to Neil Lambert for several helpful discussions during the completion of this work. I am supported by STFC studentship ST10837.

\bibliography{../Refs.bib}
\bibliographystyle{JHEP}
\bibliographystyle{plain}

\end{document}